\documentclass[aps,pra,showpacs,twocolumn,floatfix]{revtex4}

\usepackage{amsmath}
\usepackage{bm}

\begin{document}

\title{Magnetic dipole and electric quadrupole moments of the $^{229}$Th nucleus}

\author{ M. S. Safronova$^{1,2}$, U. I. Safronova$^{3,4}$, A. G. Radnaev$^{5,6}$, C. J. Campbell$^6$, and A. Kuzmich$^6$}

\affiliation {$^1$Department of Physics and Astronomy, University of Delaware, Newark, Delaware 19716\\
$^2$Joint Quantum Institute, NIST and the University of Maryland, College Park, Maryland 20742\\
$^3$Physics Department, University of Nevada, Reno, Nevada 89557,
\\$^4$Department of Physics, University of Notre Dame,
Notre Dame, IN 46556 \\
$^5$KLA-Tencor Corporation, Milpitas, California 95035\\
$^6$School of Physics, Georgia Institute of Technology, Atlanta, Georgia 30332}

\date{\today}

\begin{abstract}
We calculate the $A$ and $B$ hyperfine constants for the low-lying states of $^{229}$Th$^{3+}$ using a high-precision relativistic all-order approach. By combining these calculations with measurements of the $6d$ and $5f$ hyperfine constants [C. J. Campbell \textit{et al.}, Phys. Rev. Lett. {\bf 106}, 223001 (2011)], we determine the magnetic dipole $\mu=0.360(7)~\mu_B$ and the electric-quadrupole $Q=3.11(6)$~$e$b moments of the $^{229}$Th nucleus. Our value for $\mu$ is five times more accurate and is 22\% smaller than the best previous value $\mu=0.46(4)$~$\mu_B$ [S. Gerstenkorn \textit{et al.}, J. Phys. (Paris) 35, 483 (1974)], while our value for $Q$ is the same, but 2.5 times more accurate than the 2011 result. A systematic study of hyperfine structure in eight other monovalent atoms supports our claim of $2\%$ level uncertainty for $\mu$ and $Q$ in $^{229}$Th$^{+3}$.

\end{abstract}


\maketitle


\section{Introduction}

The $^{229}$Th  nucleus has an unusually low
first excitation energy of only several eV ~\cite{kroger,beck}, making the corresponding nuclear transition accessible with laser
excitation. The transition is expected to be very narrow and well-isolated from effects of external fields and, therefore, presents
a remarkable opportunity for the development of a nuclear clock ~\cite{peik2003}. The physical implementation may employ either the stretched states within the $5f_{5/2}$ electronic ground level of both nuclear ground and isomer manifolds of a single trapped ion ~\cite{CamRadKuz12}, or the closed electronic shell of Th$^{4+}$ in a UV-transparent crystal doped with a macroscopic number of $^{229}$Th nuclei \cite{rellergert}. Moreover, the transition frequency is expected to be as many as five orders of magnitude more sensitive to temporal variation of the fine structure constant and the dimensionless strong interaction parameter $m_q/\Lambda_{\textrm{QCD}}$ as compared to atomic transitions, making $^{229}$Th one of the most attractive candidates for such studies \cite{flambaum06}.  Th$^{3+}$ was also suggested for study of parity violation \cite{DzuFlaRob13}.

For the future development of the ion-trap nuclear clock, it is essential to possess accurate values for both nuclear and electronic properties of $^{229}$Th$^{3+}$. The electronic level structure of the Fr-like Th$^{3+}$ ion, with a single valence electron above the closed [Rn]=[Xe]$4f^{14}5d^{10}6s^26p^6$ core, is conducive to high-accuracy atomic calculations \cite{SafJohSaf06,BerDzuFla09}. Laser-cooled Wigner crystals of $^{229}$Th$^{3+}$ allow for high-precision spectroscopy  \cite{gt3,RadCamKuz12}. By combining measured hyperfine $B$ constants for the $5f_{5/2}$, $5f_{7/2}$, $6d_{3/2}$, and $6d_{5/2}$ levels with atomic structure calculations of Ref. \cite{BerDzuFla09}, the nuclear electric quadrupole moment $Q=3.11(16)$~eb was reported \cite{gt3}.

From a broader perspective, precision measurements of nuclear moments by atomic spectroscopy are essential to improve our understanding of the atomic and nuclear structure. Such measurements may also lead to an emergence of novel physics involving, e.g., breaking of a fundamental symmetry or a temporal variation of a fundamental constant.   The on-going proton radius puzzle \cite{pohl} underscores the significance of such studies. In this Letter, we report (1) a determination of the magnetic dipole and electric-quadrupole moments of the $^{229}$Th nucleus at the 2\% level of precision, and (2) a benchmark study of the accuracy of a theoretical method for the calculations of the electric quadrupole hyperfine matrix elements for future determination of electric-quadrupole moments in other systems. We also discuss the prospects for further accuracy improvements and possible extraction of the nuclear magnetic octupole moment.

The atomic hyperfine constants $A$ are proportional to the nuclear magnetic dipole moment $\mu$. Therefore, we  calculate $(A/ \mu)^{\textrm{th}}$ and determine $\mu$  as the ratio of the experimental value for the hyperfine constants $A^{\textrm{expt}}$ and $(A/ \mu)^{\textrm{th}}$. Similarly, the nuclear electric quadrupole moment $Q$ is determined as the ratio of the experimental value for the hyperfine constants $B^{\textrm{expt}}$ and the computed quantity $(B/Q)^{\textrm{th}}$.

The calculations are carried out using a relativistic linearized coupled-cluster (all-order) method including single, double, and partial triple (SDpT) excitations described in review \cite{review07} and references therein. This  method was applied to  the calculations of transition properties and polarizabilities of  Th$^{3+}$
in Refs. \cite{SafJohSaf06,SafSaf13}.
A brief description of the method is given in the supplementary material ~\cite{EPAPS}.
This method has been applied to the calculation of hyperfine constants in a number of monovalent systems \cite{SafSaf11,SafSaf11a,SimSafSaf11,Saf10,GomAubOro08,Saf10,PalJiaSaf09}.  The resulting  values for Ca$^+$, Sr$^{+}$, Hg$^{+}$,
Rb, Cs, Ba$^+$, Fr, and Ra$^+$ are compared with experimental values in Table~I of the supplementary material ~\cite{EPAPS}. The nuclear magnetic moments are taken from
\cite{Sto05} unless noted otherwise. The results for the last four systems are given in Table~\ref{comp}.
 We find that Ba$^+$ and Ra$^+$ hyperfine constants represent the best benchmark testing cases for Th$^{3+}$, therefore, we carry out new calculations for Ba$^+$ and Ra$^+$ hyperfine constants $A$ using exactly the same approach used for Th$^{3+}$.
The lowest-order values are listed in the ``DF'' column of Table~\ref{comp}. The relative correlation correction, evaluated as the difference of the \textit{ab initio} final and the lowest-order values, is listed in column ``CC''. We find excellent agreement with experimental values
with exception of the $nd_{5/2}$ states, where the correlation correction is so large that the final and lowest-order values are of opposite sign. Our Fr result for the $A(9s)$ state has been used in Ref.~\cite{GomAubOro08} to extract the value
of the $^{210}$Fr nuclear magnetic dipole moment with 1\% uncertainty.
\begin{table}
\caption{\label{comp} Comparison of the hyperfine constants $A$ (in MHz) with experiment in
$^{133}$Cs  \cite{GilWatWie83,HerHofPed85}, $^{137}$Ba$^+$ \cite{hyp-6s-82,hyp-93a,hyp-5d-86},
 $^{201}$Hg$^+$ \cite{guern}, $^{210}$Fr ($\mu=4.38(5)\mu_B$) \cite{CocThiTou85,SimZhaOro99}, $^{211}$Ra$^+$ ($\mu=0.878(4)\mu_B$)  and $^{213}$Ra$^+$ ($\mu=0.613(2)\mu_B$)
 \cite{exp_hyp1,exp_hyp2,VerGirWan10}. Theory values for Cs and Fr are from  \cite{GomAubOro08}.
Ba$^+$ and Ra$^+$ values are calculated in the present work. Lowest-order DF values
 are given in column DF and relative correlation corrections are given in column ``CC'' in \%.
$^a$Rescaled from the measurement in $^{223}$Ra.
 $^{b}$Recommended value based on Ba$^+$ data.}
\begin{ruledtabular}
\begin{tabular}{llrrrcc}
\multicolumn{1}{c}{System} &\multicolumn{1}{c}{Level} & \multicolumn{1}{c}{DF} & \multicolumn{1}{c}{Final} & \multicolumn{1}{c}{Expt.} &
\multicolumn{1}{c}{CC(\%)}&\multicolumn{1}{c}{Diff.(\%)} \\
\hline
$^{133}$Cs    &$6s      $&   1424 &   2276 &   2298     & 37&    1.0 \\
              &$7s      $&   391  &   540  &   546      & 28&    1.1 \\
              &$8s      $&   163&   216.8&   219.3(1) & 25&    1.1 \\
$^{137}$Ba$^+$&$6s      $&   2913 &   3998 &   4019     & 27&    0.5 \\
              &$6p_{1/2}$&  491 &  734.0 &  743.8     &  33 &  1.3 \\
              &$6p_{3/2}$&  71.2 &  121.3 &  127.3     & 41 & 4.7\\
              &$5d_{3/2}$&  128 &  191.5 &  189.7     &   33 &-1.0 \\
              &$5d_{5/2}$&  51.6  &  -10.0 &  -12.0     &  616 &  17 \\
$^{210}$Fr    &$7s      $&  4740  &  7244  &  7195(1)   & 35 &  -0.7   \\
              &$8s      $& 1214   & 1577   & 1578(2)    & 23  &  0.1\\
$^{211}$Ra$^+$&$7s      $& 5099   &     6728 &  6625(1)&24  &-1.6\\
              &$7p_{1/2}$& 860  &1299.1 &1299.7(8)  &   34 & 0.0  \\
$^{213}$Ra$^+$&$7s      $& 17798  & 23488  & 22920(6)  &   24 &-2.5 \\
              &$7p_{1/2}$&  3001  &  4535  &  4542(7)   &   34 & 0.2  \\
              &$7p_{3/2}$&  230   &  384$^{b}$&  384(5)$^{a}$    &   38 & 0 \\
              &$6d_{3/2}$&  360   &  538$^{b}$  &  528(6)    &   34 &-1.9
\end{tabular}
\end{ruledtabular}
\end{table}

In this work, we have carried out the calculations of the hyperfine constants for the $5f$, $6d$, $7s$, $7p$, $7d$, $8s$, and $8p$ states
using an SDpT all-order method. A Fermi distribution with the same parameters as the charged distribution is used in all calculations to model a finite magnetization distribution. The Breit interaction has been added on the same footing as the Coulomb interaction in the construction of the finite basis set. It  was found to be small, less than 1\%.

Determination of $^{229}$Th ($I=5/2$) nuclear magnetic dipole moment is illustrated in Table~\ref{tab-mu}.
  The values of $\mu$ (in $\mu_B$) are obtained as the ratio of the experimental $A$ values and
  theoretical $A/\mu$ results  for all four states. \textit{Ab initio} theory values are listed for the $5f_{5/2}$
  and $5f_{7/2}$ states. The values for $6d_{3/2}$ and $6d_{5/2}$ states are obtained by scaling \textit{ab initio} results by -0.75\% and -13.7\%, respectively,  using a comparison of theoretical $5d$ Ba$^+$ values with experimental data. The scaling factors were calculated as the difference of the value for the corresponding $5d$ Ba$^+$ constant with experiment multiplied by the ratio of the correlation corrections for the $6d_j$ state of Th$^{3+}$ and $5d_j$   state of Ba$^+$. Ba$^+$ is an excellent
  benchmark case for comparison with Th$^{3+}$ since the distribution of the correlation correction terms for all states is very similar for the two and the magnetic moment of $^{137}$Ba is known with less than 0.01\% uncertainty \cite{Sto05}. The value determined from the  $6d_{3/2}$ hyperfine constant is taken as the final result
  because of the Ba$^+$ benchmark comparison that allowed for additional improvement of the value. We note that the value extracted from the $5f_{5/2}$ state differs by only 0.38\% from the final value. The $5f_{5/2}$ \textit{ab initio} result is expected to be the most accurate since it has the smallest correlation correction, only 11\%, while the correlation correction for the $6d_{3/2}$  state is 24\%. To illustrate the consistency of the extracted $\mu$ values with all four experimental results, we
  multiply the theory values for $A/\mu$ by our final value of $\mu=0.360\mu_B$ and compare them with  experimental data ($6d_{3/2}$ will be in exact agrement since it was used to determine the final value of $\mu$). The relative  differences of theory and experiment are listed in the last column ``Diff.'' in \%.  We find that all four of the theory values are in agreement with experiment within experimental uncertainties  listed in \% in column ``Unc.'' for convinience.
 Our final value of  $\mu=0.360(7)$~$\mu_B$ is significantly lower than the previous result $\mu=0.46(4)$~$\mu_B$ \cite{Ger74} that was also derived from the hyperfine constant measurements.
  \begin{table}
\caption{\label{tab-mu} Determination of the $^{229}$Th nuclear magnetic dipole moment.   The values of $\mu$ (in $\mu_B$) are obtained as the ratios of the values in the Expt. and $A/\mu$ theory columns. The value determined from the  $6d_{3/2}$ hyperfine constant is taken as the final value. Theoretical
 values of $A$ (in MHz) with $\mu=0.360~\mu_B$ are listed in column ``A, Theory''. Relative differences of the experimental hyperfine constants and
 theoretical values with $\mu=0.360~\mu_B$ are given in column ``Diff.'' in \%. }
\begin{ruledtabular}
\begin{tabular}{lrrrcrr}
\multicolumn{1}{c}{State} &\multicolumn{1}{c}{$A$} &\multicolumn{1}{c}{Unc.}  & \multicolumn{1}{c}{$A/\mu$}
&\multicolumn{1}{c}{$\mu$} &
\multicolumn{1}{c}{$A$} & \multicolumn{1}{c}{Diff.}  \\
\multicolumn{1}{c}{}  &\multicolumn{1}{c}{Expt.}  &\multicolumn{1}{c}{Expt.} & \multicolumn{1}{c}{Theory}
&    \multicolumn{1}{c}{} &
\multicolumn{1}{c}{Theory} & \multicolumn{1}{c}{}  \\
\hline
 $5f_{5/2}$  &  82.2(6)  & 0.7\%&   229.2  &    0.359  & 82.5  &  -0.4\% \\
 $5f_{7/2}$  &  31.4(7)  & 2.2\%&   86.1   &    0.365  & 31.0  &   1.3\% \\
 $6d_{3/2}$  &  155.3(12)& 0.8\%&   431.5  &    0.360  & 155.3 &   0     \\
 $6d_{5/2}$  &  -12.6(7) &5.6\% &   -36.7  &    0.343  & -13.2 &  -4.8\% \\
 Final       &           &      &          &    0.360(7)&    &
\end{tabular}
\end{ruledtabular}
\end{table}
 \begin{table*}
\caption{\label{tab-q} Determination of the $^{229}$Th nuclear electric quadrupole moment from theory values of Ref.~\cite{BerDzuFla09}   and
present results. The ``Present work'' $B/Q$ column contains final \textit{ab initio} theory results  in MHz/eb.  The average value of $Q$
 is taken as the final value. Theoretical
 values of $B$ with $Q=3.11$~eb are listed in columns ``$B$''. The relative difference of the experimental hyperfine constants and
 theoretical values with $Q=3.11$~eb are given in column ``Diff.'' in \%.}
\begin{ruledtabular}
\begin{tabular}{lccccccccccc}
\multicolumn{1}{c}{State} &\multicolumn{1}{c}{Expt.} & \multicolumn{1}{c}{}&\multicolumn{4}{c}{Reference~\cite{BerDzuFla09}} &
 \multicolumn{1}{c}{}& \multicolumn{4}{c}{Present work} \\
\multicolumn{1}{c}{} &\multicolumn{1}{c}{$B$} & \multicolumn{1}{c}{}& \multicolumn{1}{c}{$B/Q$} & \multicolumn{1}{c}{$Q$}  & \multicolumn{1}{c}{$B$}
&\multicolumn{1}{c}{Diff.} & \multicolumn{1}{c}{} & \multicolumn{1}{c}{$B/Q$} & \multicolumn{1}{c}{$Q$}  &
\multicolumn{1}{c}{$B$}  &\multicolumn{1}{c}{Diff.}\\
\hline
$5f_{5/2}$ & 2269(6) &&740 &3.07    &2300 & -1.4\%&&725 &3.13  &  2254 &   0.7\% \\
$5f_{7/2}$ & 2550(12)&&860 &2.97    &2680 & -4.9\%&&809 &3.15  & 2515  &  1.4\%   \\
$6d_{3/2}$ & 2265(9) &&690 &3.28    &2150 & 5.3\% &&738 &3.07  &  2295 &   -1.3\% \\
$6d_{5/2}$ & 2694(7) &&860 &3.13    &2680 & 0.7\% &&873 &3.09  &  2716 &   -0.8\% \\
Final      &         &&    &3.11(16)&     &       &&    &3.11(6)&     &
\end{tabular}
\end{ruledtabular}
\end{table*}
 \begin{table}
\caption{\label{tab-all} Hyperfine constants $A$ and $B$ in
 $^{229}$Th$^{3+}$ ($I$=5/2, $\mu=0.360~\mu_B$, $Q=3.11$~eb) in MHz.
Relative  correlation corrections are listed in columns ``CC'' in \%. $^{(a)}$These three values are obtained by scaling \textit{ab initio} 156.5, -11.4, and 115.4~MHz results using Ba$^+$ data.}
\begin{ruledtabular}
\begin{tabular}{lcccclcccc}
 \multicolumn{1}{c}{Level} & \multicolumn{1}{c}{$A$} & \multicolumn{1}{c}{CC}
 & \multicolumn{1}{c}{$B$}&\multicolumn{1}{c}{CC} & \multicolumn{1}{c}{Level}&\multicolumn{1}{c}{$A$} & \multicolumn{1}{c}{CC}
 & \multicolumn{1}{c}{$B$} & \multicolumn{1}{c}{CC}\\
\hline
  $5f_{5/2}$& 82.5     & 11\%  &  2254&  26\% &  $7s$      & 5806     & 17\%  &      &        \\
  $5f_{7/2}$& 31.0     &-23\%  &  2515&  29\% &  $7p_{1/2}$& 1419     & 23\%  &      &       \\
  $6d_{3/2}$& 155.3$^a$& 24\%  &  2295&  17\% &   $8s   $   & 2114     & 15\%  &      &         \\
  $6d_{5/2}$& -13.2$^a$& 485\% &  2716&  26\% &   $8p_{1/2}$& 582      & 19\%  &      &        \\
  $7d_{3/2}$& 43.0     & 28\%  &  824 &  40\% &   $7p_{3/2}$& 119.4$^a$& 31\%  &  5310&  30\%   \\
  $7d_{5/2}$& 8.59     & -36\% &  977 &   45\%&  $8p_{3/2}$& 51.3     & 30\%  &  2238&26\%
\end{tabular}
\end{ruledtabular}
\end{table}

Determination of the $^{229}$Th nuclear electric quadrupole moment is illustrated in Table~\ref{tab-q}. The table structure is similar to that of Table~\ref{tab-mu}, but  also includes prior determination of $Q$ carried out in \cite{gt3} using theory data from \cite{BerDzuFla09} for comparison. Columns 3-6 show determination of $Q$ using theory values from \cite{BerDzuFla09}, and columns 7-10 show determination of $Q$ using our present \textit{ab initio}
 theory results. We take the average of the results obtained from four states as the final $Q$ value since the experimental and theoretical accuracy is  similar for all four states and no relevant benchmarks exist for any of the four states. The value of the Ba$^+$ nuclear quadrupole moment \cite{Sto05} relies on the old many-body perturbation theory calculations  which are of lower accuracy than the present calculation, so Ba$^+$ experimental $B$ values  can not be used as a benchmark. Remarkably, the average value of $Q$ determined using our data and Ref.~\cite{BerDzuFla09} is identical to 3 significant figures, despite 2-6\% differences in the values of $B/Q$. However, the differences of $Q$ values determined from all four $5f_j$
  and $6d_j$ states and the final value are lower in the present work, 0.7-1.3\% (compare two ``Diff''. columns).
 Our final value, $Q=3.11(6)$~eb, is significantly lower than the previous values of 4.3(9)~eb
~\cite{Ger74} also  inferred from optical hyperfine measurements, but is in agreement with the result deduced from the Coulomb excitation  of the nucleus, $Q=3.15(3)$~eb \cite{BemMcGFor88}.

 We used several different methods to establish the uncertainties in the values of $\mu$ and $Q$.

(Ia) First, we establish the size of the total correlation corrections  to the
  $5f$  and $6d$ hyperfine constants.  We list $^{229}$Th$^{3+}$ hyperfine constants $A$  and $B$  in Table~\ref{tab-all} in MHz; our final values $\mu=0.360~\mu_B$, $Q=3.11$~eb  are used.
 The relative size of the correlation corrections defined as the ratio of \textit{ab initio} (SDpT-DF)/SDpT, where DF are the lowest-order results, is listed in columns ``CC'' in \%.  We find that the correlation corrections to the $6d$ and $5f$ $A$ and $B$ values are relatively small, 11-29\%, with the only exception of $A(6d_{5/2})$, where the correlation correction is so large that
 the lowest-order DF value has a sign opposite that of the final value.  The uncertainty of both theory and experiment is about 5\% for the $6d_{5/2}$ state, so it was not used in determination of $\mu$.

 (Ib) Next, we establish the clear correlation between the size of the correlation corrections and the accuracy of the values using the comparison of the theory and experimental values for Ca$^+$, Sr$^{+}$, Hg$^{+}$,
Rb, Cs, Ba$^+$, Fr, and Ra$^+$,  given in Table~I of the supplementary material ~\cite{EPAPS}.
All cases similar to the $5f$ and $6d$ states of Th$^{3+}$ where the correlation correction is below 30\% agree with experiment to better than 2\%. We note that Hg$^+$ is not a true ``monovalent'' case owing to low-lying core excitations that are not present in Th$^{3+}$, so agreement is slightly worse (-2.3\%). We find the average ratio of the last two columns of Table~I of ~\cite{EPAPS} (that give a correlation correction and difference with experiment) is 2.9\%, and the maximum ratio is 6\% [we excluded Hg$^+$ and the anomalous case of the $4d_{3/2}$ state in Rb, where the correlation is 65\% for the average]. If we average over only heavy systems listed in Table~\ref{comp} (Cs, Ba$^+$, Fr, Ra$^+$) the average ratio is still 3.1\%.
Therefore, we can make a general conclusion that the uncertainty of the calculations is expected to be on the order of 3\% of the total correlation correction  and should not exceed 6\%. For the $6d_{3/2}$ and $5f_j$ $A$ constants,  3\%/6\% of the correlation gives (0.3-0.7\%)/(0.6-1.4\%) uncertainty, depending on the state, with $5f_{5/2}$ being the most accurate.
For all $6d_j$ and $5f_j$ $B$ constants,  3\%/6\% of the correlation gives (0.5-0.9\%)/(1-1.8\%) uncertainty, depending on the state, with $6d_{3/2}$ being the most accurate. This analysis gives 1.5\% upper bound on the theory uncertainty for $\mu$ and 2\% for $Q$. While the above analysis is done for $A$ constants, it holds for a variety of states with completely different distributions of the correlation correction terms. Therefore, it is expected to be applicable to $B$ constants as well.
The experimental uncertainty is 0.7\% for $A(6d_{3/2})$ and $A(5f_{5/2})$, giving a 2\% combined upper bound on the uncertainty of $\mu$. The experimental uncertainty for all $B$ constants is 0.26-0.47\%, which can be considered negligible in comparison with the theory uncertainty.

 (II) We further compare a benchmark case of $A(6d_{3/2})$ with $A(nd_{3/2})$ Ba$^+$ and Ra$^+$ measurements since our final determination of $\mu$ is based on this value.
 To ascertain that these cases are indeed very similar, we compare the entire breakdown of the 20 correlation
    correction terms and normalization contributions to the hyperfine constants. The comparative
    breakdown of correlation terms  for various  states
    of Ba$^+$, Ra$^+$, and Th$^{3+}$ is given in Table II of the supplementary material~\cite{EPAPS}. We find that the Ba$+$ and Ra$^+$ $nd_{3/2}$ cases are nearly identical, while the Th$^{3+}$ $6d_{3/2}$ case is very similar, but has a smaller overall correlation correction, 24\%, instead of 33-34\%. This is expected as the correlation effects decrease with increasing degree of ionization.
As we noted above, we use a $-$1\% difference of our Ba$^+$ value with experiment and a 24/34 correlation correction ratio
to adjust our $6d_{3/2}$ value by $-$0.75\%. As a result, we expect that our $6d_{3/2}$ theory value is accurate to better than 1\%. However, the difference for the similarly adjusted $6d_{3/2}$ hyperfine constant of $^{213}$Ra$^+$ with experiment is $-$1.9\%. The Ra$^+$ measurement uncertainty is 1.1\% for this state, and the uncertainty of the $^{213}$Ra$^+$ magnetic moment (from a direct measurement)  is 0.3\%. We note that there is a 1\% inconsistency in the values of measured $7s$ hyperfine constants in $^{211}$Ra$^+$ and $^{213}$Ra$^+$ isotopes and our corrected (using Ba$^+$ data) values of the $6p_{3/2}$ constant is in perfect agreement with the experiment. Therefore, it is likely that the discrepancy is due to uncertainty in the Ra$^+$ measurement. Nevertheless, we keep our original 2\% estimated uncertainty of $\mu$.

(III) Finally, we use the consistency of $\mu$ and $Q$ obtained from different states as an independent uncertainty estimate.
Table II of the supplementary material~\cite{EPAPS} illustrates that the contributions of the various correlation terms  are quite different for the $6d_{3/2}$, $5f_{5/2}$, and $5f_{7/2}$ states. In fact, the correlation correction is negative for the $5f_{7/2}$ level while it is positive for the other two levels.  Therefore, these calculations are sufficiently different that the spread of the $\mu$ values obtained from three levels (0.8-1.3\%) provides another independent estimate of the accuracy. The 2\% uncertainty on the $A(5f_{7/2})$ measurement does not presently allow us to significantly reduce the uncertainty. The comparative breakdown of the correlation correction to the $B$ constants for the  $6d$ and $5f$ levels given in Table~III of the supplementary material
also shows differences between the correlation terms for the $6d$ and $5f$ levels.  The 0.7-1.4\% difference of the $Q$ values obtained for four levels with the average final value confirms a 2\% uncertainty estimate for $Q$.

The present accuracy of the value of $\mu$ can be improved (to 1\% uncertainty or better) by remeasuring $A$
for the $6d_{3/2}$ and $5f_{j}$ levels to 0.3\% or better. If $\mu$ derived from all three states were found to be consistent to 0.5-1\% with improved experimental accuracy, this would demonstrate higher accuracy of the theory values and, subsequently,  $\mu$. The measurements of the $7s$ and $7p_{1/2}$ hyperfine constants $A$ can be used to better understand finite magnetization distribution effects, serve as an independent benchmark test case  for Fr and Ra$^+$ measurements, and may provide improvement in the accuracy of $\mu$. Our calculations give 3.1\% and 1.1\% finite magnetization distribution corrections for the $7s$ and $7p_{1/2}$
states which may be a substantial overestimation based on rough nuclear calculations ~\cite{GomAubOro08}. This seems to be confirmed by differences in comparison of Cs/Fr and Ba$^+$/Ra$^+$ values for these states  as finite magnetization distribution effects  are much smaller in Cs and Ba$^+$. While the uncertainty of our \textit{ab initio} calculations for $A(7p_{3/2})$ is about 3-4\%, our scaled value should be accurate to about 1\%. The magnetization distribution effects are essentially zero for all states other than the $7s$ and $7p_{1/2}$ states.

Improved measurements of the $6d_{3/2}$ and $5f_{j}$ hyperfine intervals are critical for further improvements of the nuclear magnetic dipole moment value.  Using an ultracold sample sympathetically cooled with, e.g., $^{232}$Th$^{3+}$, and sub-kHz linewidth lasers, the $\sim$ 100 kHz wide optical transitions of interest could all be measured with inaccuracies of $<10$ kHz.  In extracting the hyperfine intervals, this would reduce the current uncertainties in the $A$ constants from $\sim 10^{-2}$ to $\sim 10^{-5}$, a level well below that of the companion calculations.  Such measurements would also reduce the current uncertainties in the measured $B$ constants by 3 orders of magnitude.

The structure of the $^{229}$Th$^{3+}$ nucleus may also be studied beyond the first two electromagnetic moments.  The 5$f_{5/2}$ electronic ground level is well suited for such studies, as it has a large total angular momentum.  Because $J$ is greater than 1, the valence electron has non-zero coupling to the nuclear magnetic octupole moment $\Pi$.  This effect, parameterized by the magnetic octupole hyperfine constant $C$, is expected to shift hyperfine intervals at the level of $10^1-10^2$ Hz.  Observing such an effect is straightforward in this ground-state hyperfine manifold, using microwave spectroscopy, as the individual states are effectively infinitely narrow and each hyperfine level contains an $m_F = 0$ clock state, allowing for the removal of first-order Zeeman shifts from the measurements. All hyperfine intervals should be measurable with inaccuracies of $\sim$ 0.01 Hz.  This level of error is $<1 \%$ of an expected minimum value for $C$, indicating that percent-level inaccuracies in atomic structure calculations and nuclear magnetic dipole and electric quadrupole moment values would lead to extraction of $\Pi$ with an uncertainty of at most a few percent, on par with the most accurately determined nuclear octupole moment to date \cite{MDBarrett2012}.

In conclusion, we determined the nuclear magnetic dipole and electric quadrupole moments of $^{229}$Th with 2\% uncertainty by combining high-precision theoretical and experimental values of the hyperfine constants and outline a pathway for further improvements in the accuracy of these properties.  The present work also establishes a high-precision benchmark for accurate determination of electric quadrupole  nuclear moments in other systems.

The work of M.S.S.
was supported in part by National Science Foundation Grant No.\ PHY-1068699, work of A.R., C.C., and A.K. was supported by the National Science Foundation and the Office of Naval Research.


\end{document}